# The "Egg of Columbus" for making the world's toughest fibres

Nicola M. Pugno[1,2]

[1]Laboratory of Bio-Inspired & Graphene Nanomechanis,
Department of Civil, Environmental and Mechanical Engineering,
Università di Trento, via Mesiano, 77, I-38123 Trento, Italy,
[2]Center for Materials and Microsystems, Fondazione Bruno Kessler,
Via Sommarive 18, I-38123 Povo (Trento), Italy.
nicola.pugno@unitn.it,

**A great flourish of interest in the development of new high-strength and high-toughness materials is taking place in contemporary materials science, with the aim of surpassing the mechanical properties of commercial high-performance fibres. Recently, macroscopic buckypapers[1-5], nanotube bundles[5-13] and graphene sheets[14-17] have been manufactured. While their macroscopic strength remains 1-2 orders of magnitude lower than their theoretical strength, and is thus comparable to that of current commercial fibres, recent progress has been made in significantly increasing toughness. In particular, researchers have produced extremely tough nanotube fibres with toughness modulus values of up to 570 J/g[8], 870 J/g[13] and very recently, including graphene, reaching 970 J/g[18], thus well surpassing that of spider silk (~170 J/g[8], with a record for a giant riverine orb spider of ~390 J/g[19] and Kevlar (~80 J/g[8]). In this letter, thanks to a new paradigm based on structural mechanics rather than on materials science, we present the "Egg of Columbus" for making fibres with unprecedented toughness: a slider, in the simplest form just a knot, is introduced as smart frictional element to dissipate energy and in general to reshape the fibre constitutive law, showing evidence of a previously "hidden" toughness, strictly related to the specific strength of the material. The result is a nearly perfectly plastic constitutive law, with a shape mimicking that of spider silk. The proof of concept is experimentally realized making the world's toughest fibre, increasing the toughness modulus of a commercial Endumax fibre from 44 J/g up to 1070 J/g. The maximal achievable toughness is expected for graphene, with an ideal value of ~$10^5$ J/g.**

The *concept* is thus based on the introduction of appropriate sliders, even simple knots, as smart frictional elements for energy dissipation in high-strength fibres. Part of the energy is dissipated through friction by the fibre sliding in the slider during tension, additionally to the intrinsic stretching energy dissipated by fracture and kinetic energy when the fibre breaks.

The energy at break $\phi$ per unit mass $m$ of a fibre of cross-sectional area $A$, length $l$, Young's modulus $E$, strength $\sigma_f$ and mass density $\rho$, can be calculated from the load-displacement or stress-strain curve as $\phi/m = 1/m \int_0^{x_f} F dx = l_0 A/m \int_0^{\varepsilon_f} \sigma d\varepsilon = (1-k_1)/\rho \int_0^{\varepsilon_f} \sigma d\varepsilon$, where $F$ is the force, $x$ is the displacement, $\sigma = F/A$ is the stress, $\varepsilon = x/l_0$ is the strain, $l_0$ is the end-to-end length of the fibre (see Fig.1), $0 \leq k_1 = (l-l_0)/l \leq 1$ is a slider parameter and the subscript $f$ denotes final values. For a linear elastic fibre, classical thus without a slider ($k_1 = 0$), this simply yields $\phi/m = \sigma_f \varepsilon_f /(2\rho)$, where $\varepsilon_f = \sigma_f/E$. Let us now consider a fibre forming a large slip-loop in a slider, e.g. a knot, with the two clamped fibre ends initially fixed as close as possible to the slider, Fig. 1. In this case, after fibre tension, first the strain increases with the fibre sliding through the slider at a mean stress plateau value of $\sigma_p \approx k_3 \sigma_k$, where $\sigma_k = k_2 \sigma_f$ is the strength of the fibre in the presence of the slider and $0 \leq k_{2,3} \leq 1$ denote the ratios between slider and fibre strengths and between plateau stress and slider fibre strength, respectively. This sliding phase takes place for a displacement $\Delta x \approx l - l_0$, ideally at a force just below the breaking force $F_f$ of the fibre, in order to maximize the dissipated energy; then, the slip-loop tightens (it could also unfasten, depending on the type of slider/knot topology, e.g. see Fig. 2) and the fibre deforms and finally breaks. Thus, in the stress-strain curve a long plastic-like plateau naturally emerges (Figs. 2 and 3) thanks to the presence of the slider. The increment of the final strain is $\Delta \varepsilon_f \approx \Delta x/l_0 = k_1/(1-k_1)$ and can be precisely tuned selecting an appropriate value of $k_1$. The strength of the fibre with the slider (e.g. "knot strength") should ideally approach that of the pristine fibre in order to minimize the negative strength variation $\Delta \sigma_f = \sigma_f - \sigma_k = (k_2 - 1)\sigma_f$. Accordingly, the increment in toughness modulus is given by $\Delta \phi/m \approx k_1 k_2 k_3 \sigma_f / \rho$. Thus, the slider provides a potential toughness reaching the enormous value of $\sigma_f/\rho$, which is the fibre specific strength. Regarding the estimation of the Young's modulus $E_k$ of the knotted fibre, we simply (series of compliances) find $E_k \approx E/(1 + c_k EA/l_0)$, where $c_k$ is the knot compliance and thus a softening $\Delta E = E - E_k = (1-k)E$ ($k \approx c_k EA/l_0$ for $c_k \to 0$, i.e. tight knots, for which $E_k \to E$) is expected. More in general, the application of this concept allow to reshape the constitutive law of the fibre in terms of variations of the four main engineering properties $\Delta \phi/m, \Delta \varepsilon_f, \Delta \sigma_f, \Delta E$, i.e. toughness modulus, failure strain, strength and Young's modulus, respectively. In particular, we can mimic the constitutive law of spider silk, crucial for structural robustness[20], with the friction in the slider that plays the role of the hydrogen bond breaking in the silk, which is responsible for its dissipative plastic-like plateau. Moreover, the usually competing material properties of strength and toughness (we are considering here the toughness modulus rather than so-called fracture toughness) are reconciled: high strength and simultaneously super-tough fibres become feasible, of course at the expense of a larger elongation, the latter being either positive or negative depending on the specific application.

It is clear that in order to maximize the dissipated energy and therefore the toughness modulus, the specific strength of the fibre must be high, i.e., $\sigma_f/\rho \to \max$ (fibre condition n. 0), and the following 3 conditions must be met: the sliding length must be maximized ($k_1 \to 1$, geometrical condition n. 1); the fibre must display minimal fragilization due to the presence of the slider (expected as a consequence of the applied additional stresses; $k_2 \to 1$, fibre-slider condition n. 2); the slider must ensure an initial stress plateau as flat and as high as possible ($k_3 \to 1$, slider condition n. 3). Clearly, the mass of the slider should also be minimized, but its influence on the toughness modulus approaches zero for increasing sliding/fiber length. These 1+3 main conditions can substantially be achieved through careful choice of fibres and sliders. For example hundreds of knots, widely used in climbing, sailing and fishing activities, are currently known, and multiple knots can be realized too, allowing the design of complex, e.g. multi-plateau, constitutive laws even in complex architectures, e.g. ropes, fabrics, hierarchical reinforcements for composites, etc. Moreover, note that this concept partially applies also to other systems not in tension and without knots, e.g. curved wires under bending or torsion in addition to stretching: the wire will exhibit an increasing stiffness with deformation and the resulting toughness will be much superior than that of the related initially straight wire; eventually, the wire will straighten and fracture at the same load of an initially straight wire; however note that, in this case, the force is not maximal from the beginning and thus the gain in toughness is not maximized, as for the case of fibres working just below the fracture force from the beginning; formation of plastic hinges would be a limitation and thus super-elastic, e.g. NiTi, alloys, must be preferred. Coiled structures are examples of such smart elements, so abundant in Nature, e.g. the coiled coil structural motif of proteins (in which alpha-helices are coiled together like the strands of a rope; dimers and trimers are the most common types) but also of great interest in man-made systems such as springs, however currently not used as tougher wires.

The *proof of concept* is realized considering commercial Dyneema fibres (fishing line, multifilament composed by 200 filaments), which combine a high-strength with low density, i.e. quite high specific strength (and stiffness). We tested them in a uniaxial loading tensile testing machine (MTS, Insight). The measured constitute law of this fibre is reported in Fig. 2 (curve "without knot") and shows a specific strength of $\sigma_f/\rho \approx 1\,\text{GPa}/(1\,\text{g/cm}^3) = 1000\,\text{J/g}$ and a toughness modulus of 27 J/g. Note that the value of the specific strength, which according to our previous considerations also denotes the hidden toughness, suggests a huge margin for increasing the toughness of this fibre. Additionally, these fibres display reduced fragilization when tied into slip-knots, allowing us to prove the concept with a simple knot, that however is itself sliding along the fibre and is not an independent slider and thus does not allow us to reach a high stress plateau: here we have deliberately not optimized the knot in order to prove the robustness of the concept (in our cases $k_3 \approx 0.1 - 0.4$, depending on the number of coils, and $k_1 \approx 0.75 - 1$). The adopted knots were simple, double, triple or quadruple overhand knots (Fig. 2, inset). Different numbers of coils ($1-4$) are considered in order to maximize the friction force during sliding without leading to premature fibre fracture before full knot tightening. The fibre length is $l$=10cm whereas the selected initial value of the end-to-end length was fixed at $l_0 = 1\,\text{cm}$, thus $k_2 = 0.9$ and $\Delta\varepsilon_f = 900\%$. The measured specific stress-strain curves of knotted fibres are shown in Fig. 2 and compared to that of the unknotted fibre. The expected appearance with the knot of a previously hidden toughness, and thus of a, even if here irregular, plastic-like plateau, absent in the constitutive law of the unknotted fibre, is evident. For 1 and 2 coils a smart mechanism is observed, leading to the optimal condition $k_2 = 1$ and no strength reduction, i.e. $\Delta\sigma_f = 0$: the knot unfastens (this is the reason for which the stress goes to zero in the related curves), then the fibre extends, deforms and fractures at exactly the pristine fibre strength. For 1 coil the toughness modulus reaches 88 J/g whereas for 2 coils a value of 195 J/g is obtained, thanks to an increment in the toughness up to 722%, without any strength reduction. The constitutive law of this fibre resembles that of spider silk (in terms of both toughness modulus and strength). For 3 coils the dissipated energy further increases up to a value of 320 J/g, corresponding to a toughness increment of 1185%, with a strength reduction of ~25%. For 4 coils premature failure leads to a reduction in both failure strain and toughness modulus, of 148 J/g, with a similar strength reduction. We conclude that a number of 3 coils maximizes the toughness of our specific knot-fibre system. These set of experiments prove the robustness of the concept.

We finally consider independent sliders/knots and Endumax fibres. Following the previous procedure a more regular sliding, thanks to the independency of the slider that thus remains fixed during the fibre sliding (specific details of the slider are under patenting) is achieved and results in a more regular plastic-like plateau, see Fig. 3. The initial toughness modulus of the pristine fibre thus without slider is 44 J/g. With the introduction of the slider we obtained 988, 1025 and finally 1070 J/g, thus surpassing the unprecedented value of 1000 J/g. Further improvements are expected for this specific fibre, up to a toughness modulus close to the fibre specific strength that we have measured as 1600 J/g (see stress peak in Fig. 3).

Accordingly, thanks to the new evidence of existence of a previously "hidden" toughness, much higher toughness levels, which can surpass the highest-known values in the literature, could be obtained by applying our concept to other high-strength or super-tough fibres, e.g. graphene/nanotube-based (which have also shown to be resilient to knots[21]). Note that, considering the theoretical nanotube/graphene strength a toughness limit of about $\sigma_{theo}/\rho \sim 10^5$ J/g is computed; moreover, since smaller is stronger, the hidden toughness can be more easily observed reducing the system size-scale. In spite of this evidence of huge hidden toughness in artificial fibers, some spiders, such as the giant riverine orb spider[19], seem to have better optimized toughness thanks to the last ~200 millions of years of evolution; e.g., for the mentioned spider we compute a toughness limit of $\sigma_{theo}/\rho \sim 1380$ J/g against its measured value and actual record of ~390 J/g (we have assumed a mean density of ~1.34 g/cm$^3$). Thus, even this spider (or us, using its silk and our concept) could further improve the toughness of its silk by a factor of ~350%. During the next millions of years this percentage of hidden toughness will be probably further reduced by evolution and the same appearance of knots/sliders/current "egg of Columbus" cannot be excluded. We can do better in the next few years.

NMP is supported by the European Research Council (ERC StG 2011 BIHSNAM on "Bio-inspired hierarchical super-nanomaterials"; one milestone of this project is the design and fabrication of the "world's toughest material", as we have achieved in this paper. NMP thanks E. Lepore and F. Bosia for the technical experimental help.

This paper was submitted to a major journal; one referee was very positive writing "This manuscript described a very simple process, tying slip knows in fibers, which can significantly increase the toughness of the resultant fibres. Like many good ideas, this one is very simple. It lead me to think - why didn't I think of that. I think it should be published." and I thank this referee. The other referee suggested rejection writing sentences promoting plagiarism like "this referee would be tempted to write an equivalent paper" and "This reviewer could then proceed and test superelastic NiTi alloys and show this effect in a most eloquent manner." and the paper was rejected.

1. R. H. Baughman, C. Cui, A. A. Zakhidov, Z. Iqbal, J. N. Barisci, G. M. Spinks, G. G. Wallace, A. Mazzoldi, D. D. Rossi, A. G. Rinzler, O. Jaschinski, S. Roth, M. Kertesz, Carbon Nanotube Actuators, *Science*,1999, 284, 1340–1344.

2. Z. Wu, Z. Chen, X. Du, J. M. Logan, J. Sippel, M. Nikolou, K. Kamaras, J. R. Reynolds, D. B. Tanner, A. F. Hebard, A. G. Rinzler, Transparent conductive carbon nanotube films,*Science*,2004, 305, 1273–1273.

3. M. Endo, H. Muramatsu, T. Hayashi, Y. A. Kim, M. Terrones, M. S. Dresselhaus, Nanotechnology: 'Buckypaper' from coaxial nanotubes, *Nature*, 2005, 433, 476–476.

4. S. Wang, Z. Liang, B. Wang, C. Zhang, High-strength and multifunctional macroscopic fabric of single-walled carbon nanotubes, *Advanced Materials*2007, 19, 1257–1261.

5. M. Zhang, S. Fang, A. A. Zakhidov, S. B. Lee, A. E. Aliev, C. D. Williams, K. R. Atkinson, R. H. Baughman, Strong, transparent, multifunctional, carbon nanotube sheets, *Science* 2005, 309, 1215–1219.

6. H. W. Zhu, C. L. Xu, D. H. Wu, B. Q. Wei, R. Vajtai, P. M. Ajayan, Direct synthesis of long single-walled carbon nanotube strands, *Science* 2002, 296, 884–886.

7. K. Jiang, Q. Li, S. Fan, Nanotechnology: Spinning continuous carbon nanotube, *Nature* 2002, 419, 801–801.

8. A. B. Dalton, S. Collins, E. Munoz, J. M. Razal, Von H. Ebron, J. P. Ferraris, J. N. Coleman, B. G. Kim, R. H. Baughman, Super-tough carbon-nanotube fibres, *Nature* 2003, 423, 703–703.

9. L. M. Ericson, H. Fan, H. Peng, V. A. Davis, W. Zhou, J. Sulpizio, Y. Wang, R. Booker, J. Vavro, C. Guthy, A. N. G. Parra-Vasquez, M. J. Kim, S. Ramesh, R. K. Saini, C. Kittrell, G. Lavin, H. Schmidt, W. W. Adams, W. E. Billups, M. Pasquali, W.-F. Hwang, R. H. Hauge, J. E. Fischer, R. E. Smalley, Macroscopic, neat, single-walled carbon nanotube fibers, *Science* 2004, 305, 1447–1450.

10. M. Zhang, K. R. Atkinson, R. H. Baughman, Multifunctional carbon nanotube yarns by downsizing an ancient technology, *Science* 2004, 306, 1358-–1361.

11. Y.-L. Li, I. A. Kinloch, A. H. Windle, Direct spinning of carbon nanotube fibers from chemical vapor deposition synthesis, *Science* 2004, 304, 276–278.

12. K. Koziol, J. Vilatela, A. Moisala, M. Motta, P. Cunniff, M. Sennett, A. Windle, High-performance carbon nanotube fiber, *Science* 2007, 318, 1892–1895.

13. P. Miaudet, S. Badaire, M. Maugey, A. Derr, V. Pichot, P. Launois, P. Poulin, and C. Zakri, Hot-drawing of single and multiwall carbon nanotube fibers for high toughness and alignment,*Nano Letters 2005,* 5, 2212–2215.

14. K. S. Novoselov, A. K. Geim, S. V. Morozov, D. Jiang, Y. Zhang, S. V. Dubonos, I. V. Grigorieva, A. A. Firsov, Electric field effect in atomically thin carbon films, *Science* 2004, 306, 666–669.

15. C. Berger, Z. Song, X. Li, X. Wu, N. Brown, C. Naud, D. Mayou, T. Li, J. Hass, A. N. Marchenkov, E. H. Conrad, P. N. First, W. A. de Heer, Electronic confinement and coherence in patterned epitaxial graphene, *Science* 2006, 312**,** 1191–1196.

16. S. Stankovich, D. A. Dikin, G. H. B. Dommett, K. M. Kohlhaas, E. J. Zimney, E. A. Stach, R. D. Piner, S. T. Nguyen, R. S. Ruoff, Graphene-based composite materials, *Nature* 2006, 442**,** 282–285.

17. D. A. Dikin, S. Stankovich, E. J. Zimney, R. D. Piner, G. H. B. Dommett, G. Evmenenko, S. T. Nguyen, R. S. Ruoff, Preparation and characterization of graphene oxide paper, *Nature* 2007,448**,** 457–460.

18. M. K. Shin, B. Lee, S. H. Kim, J. A. Lee, G. M. Spinks, S. Gambhir, G. G. Wallace, M. E. Kozlov, R. H. Baughman, S. J. Kim, Synergistic toughening of composite fibres by self-alignment of reduced graphene oxide and carbon nanotubes, *Nature Comm*, 2012, 3, 650 (8 pp).

19. I. Agnarsson, M. Kuntner, T. A. Blackledge, Bioprospecting finds the toughest biological material: extraordinary silk from a giant riverine orb spider, *PLoS One*, 2010, 5 (8 pp.).

20. S.W. Cranford, A. Tarakanova, N. Pugno, M.J. Buehler, Nonlinear material behaviour of spider silk yields robust webs, *Nature*, 2012, 482, 72–78.

21. J. J. Vilatela, A. H. Windle, Yarn-Like Carbon Nanotube Fibers, *Advanced Materials 2010,* 22, 4959–4963.

22. N. Pugno, The design of self-collapsed super-strong nanotube bundles, *J. of the Mechanics and Physics of Solids*, 2010, 58, 1397–1410.

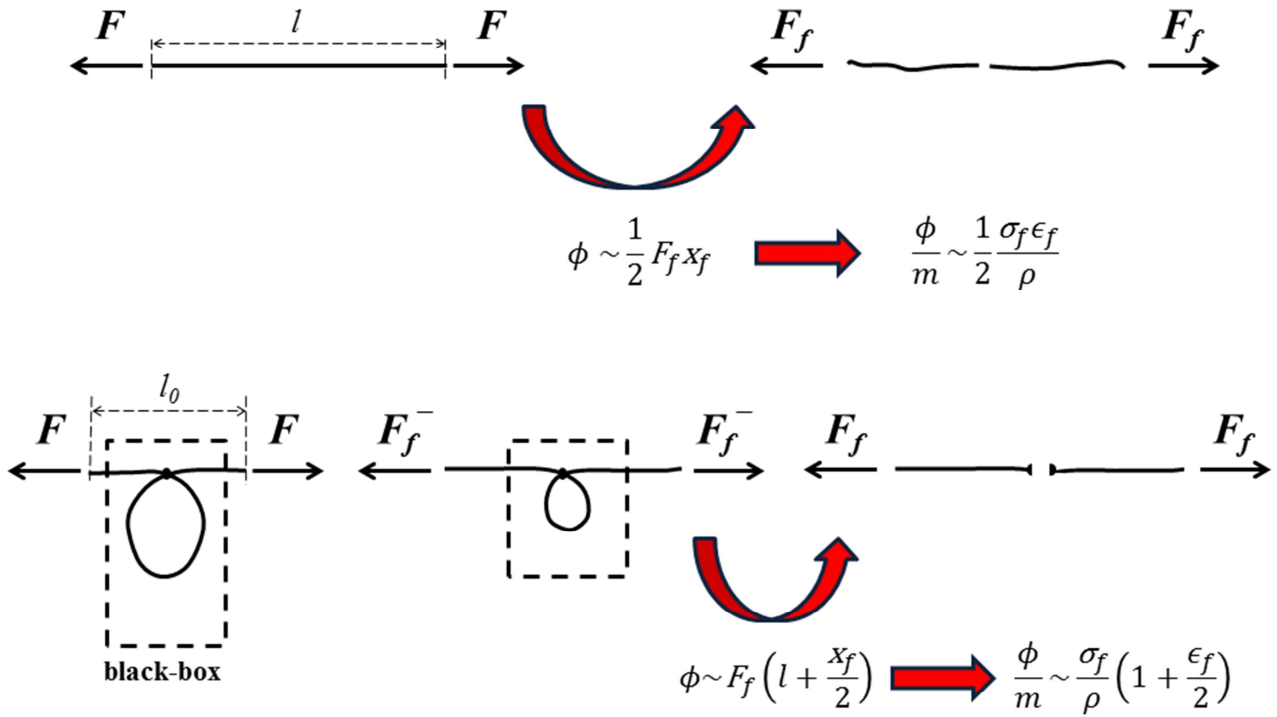

*Figure 1: Concept. The classical fibre dissipates during fracture its cumulated strain energy, thus displaying a toughness modulus of $\phi/m \approx \sigma_f \varepsilon_f /(2\rho)$ (the factor 2 must be replaced for nonlinear elastic fibres). In contrast, a fibre with a slider, e.g. knot, can dissipate much more energy, thanks to a sliding friction force. The upper limit of the toughness in this case is constituted by the product of a force $F_f^-$ just below the breaking force $F_f$ and a displacement equal to the entire fibre length l, thus reaching a toughness modulus of $\phi/m \approx \sigma_f /\rho (1 + \varepsilon_f /2)$. Accordingly, a huge ($1 \gg \varepsilon_f$) previously "hidden" toughness $\Delta\phi/m \approx \sigma_f /\rho$ naturally emerges.*

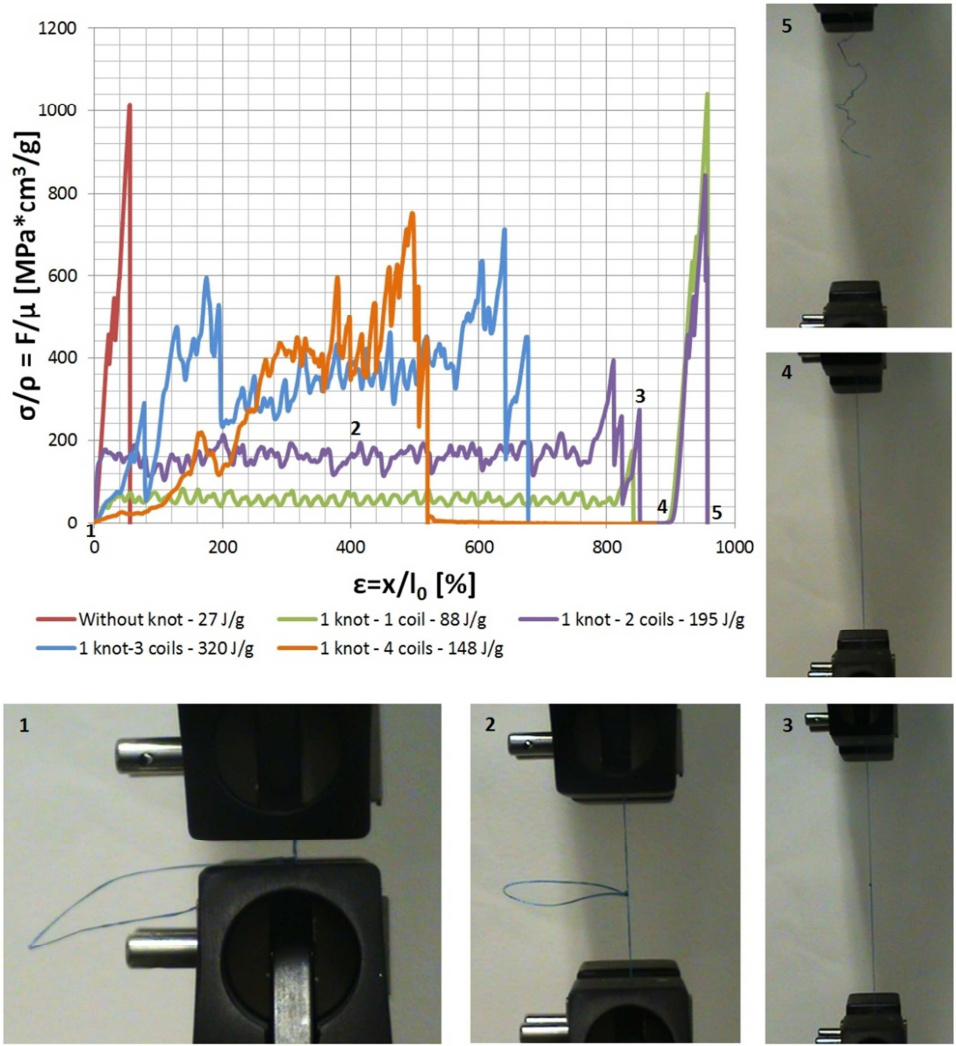

*Figure 2: Proof of concept. Specific force-displacement or stress-strain curves of non-knotted and knotted Dyneema fibres ($\sigma/\rho = F/\mu$ with $\mu = \rho A$ mass per unit length, given in MPa*cm$^3$/g or, equivalently, in J/g; test parameters are $dx/dt=2$mm/min, $l_0=10$ mm, $l=100$ mm, $\mu=0.0361$ g/m). The appearance with the knot of the hidden toughness, the plastic-like plateau absent in the constitutive law of the unknotted fibre (27 J/g), is evident. For 1 and 2 coils the knot unties (smart mechanism) and stress goes to zero, then the fibres extends, deforms and fractures at the pristine fibre strength, with an increment in the toughness of up to 722% (2 coils, 195 J/g). For 3 coils the dissipated energy is further increased up to a maximal value of 320 J/g, corresponding to a toughness increment of 1185%. For 4 coils, premature failure leads to a reduction in both toughness and failure strain. The total hidden toughness is given by the specific strength, thus for this fibre it is close to 1000 J/g.*

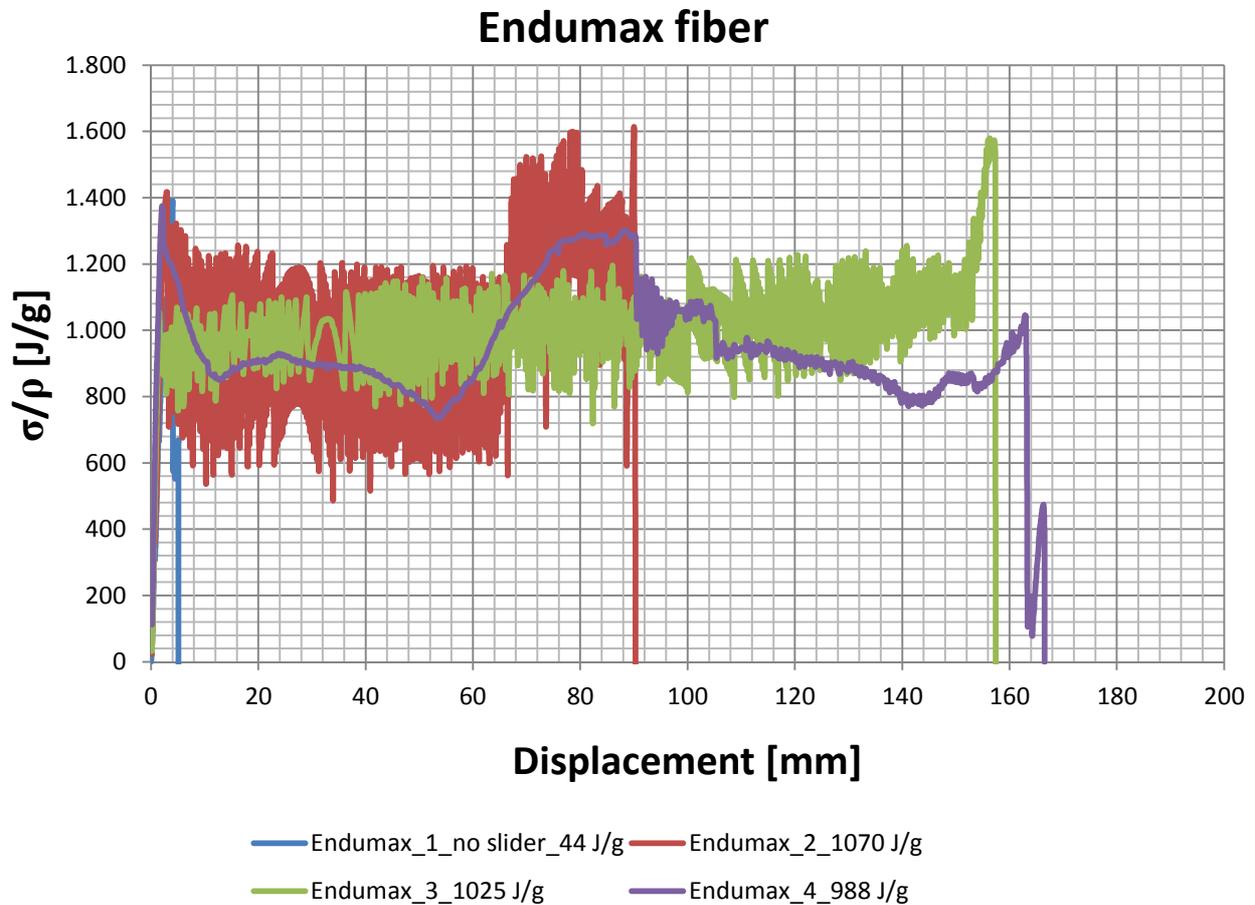

*Figure 3: World toughness record (work in progress). Force vs displacement curve of Endumax fibres with or without a slider. The pristine fibre has a toughness modulus of 44 J/g. The introduction of the slider dramatically changes the scenario: a long plastic-like plateau clearly emerges thanks the presence of the slider and allows the dissipation of a huge amount of energy, approaching an unprecedented toughness modulus of 1070 J/g (other two tests, leading to 988 and 1025 J/g are shown). The specific strength and thus maximal achievable toughness is for this fibre of about 1600 J/g (see stress peak in the figure).*